\documentstyle[preprint,aps]{revtex}
\begin{document}
\tightenlines
\title{The propensity of molecules to spatially align in intense light fields} 
\author{S. Banerjee, D. Mathur, and G. Ravindra Kumar}
\address{Tata Institute of Fundamental Research, Homi Bhabha
Road, Mumbai 400 005, India}
\date{\today}
\maketitle
\begin{abstract}
The propensity of molecules to spatially align along the polarization vector 
of intense, pulsed light fields is related to readily-accessible 
parameters (molecular polarizabilities, moment of inertia, peak intensity of 
the light and its pulse duration). Predictions can now be made of which 
molecules can be spatially aligned, and under what circumstances, upon 
irradiation by intense light.  Accounting for both enhanced ionization and 
hyperpolarizability, it  is shown that {\it all} molecules can be 
aligned, even those with the smallest static polarizability, when 
subjected to the shortest available laser pulses (of sufficient intensity).  
\end{abstract}
Studies of the response of matter to very intense fluxes of electromagnetic radiation 
address fundamental physics issues of systems driven strongly away from equilibrium. 
Matter is inherently unstable when subjected to strong electric fields of the type that 
can be generated in intense laser light. In the case of molecules subjected to laser 
light of intensity in excess of $\sim$10$^{12}$ W cm$^{-2}$, distortions of 
potential energy surfaces, with concomitant alterations in electron density distributions,  
lead to ionization, dissociation, and formation of strong dipole moments 
($\bf{\mu}$). With linearly polarized ${\bf{E}}$ fields of magnitudes that equal, 
or exceed, interatomic binding fields, the induced dipole moments exert torques on 
the molecular axes, ${\bf{\mu}\times\bf{E}}$, that can be large enough to spatially 
reorientate molecules and their ions such that the most polarizable molecular axis 
points along the light field vector. In early experiments,  anisotropic angular 
distribution of fragment ions were 
obtained when the light polarization vector was rotated relative to the detector axis, 
and these were taken to be unambiguous signatures of spatial orientation 
\cite{angular1,angular2}. In recent studies, it has been recognized that the molecular 
ionization 
rate depends on the angle that the internuclear axis makes with the light field vector, 
and that this also leads to anisotropic angular distributions \cite{angular3}. Indeed, 
the ionization rate has to be computed using the field ionization Coulomb explosion 
model \cite{ljf1}, whose angular dependence arises from the fact that the barrier 
suppression is given by $\bf{E}\cdot\bf{r}$,  where $\bf{r}$ denotes the 
molecular axis. 

Spatial alignment of isolated molecules is a subset of one of the central endeavors 
of physicists and chemists, namely to control the external degrees of freedom of 
atomic and molecular species at the microscopic level. The polarizability interaction 
of an intense, linearly-polarized light field with the induced dipole moment of 
molecules gives rise to a double-well potential; the resulting angular reorientation 
of molecular axes is akin to the interconversion of left- and right-handed enantiomers 
that was considered by Hund over 3 decades ago in terms of similar potentials 
\cite{hund}. Spatial alignment of individual molecules can also justifiably be 
considered a special facet of the optical Kerr effect 
\cite{williams}.  On a more practical level, interest in studies of spatial alignment 
of molecules has been generated because of tantalizing possibilities of entirely new  
studies on pendular-state spectroscopy \cite{pendular}, coherent control experiments 
\cite{charron}, and molecular trapping and focusing \cite{trapping}. The crucial role
of polarization in choosing or altering dissociation pathways has also been
experimentally established \cite{ch4}. It is clearly very important, 
therefore, to establish, both on the basis of the properties of the molecule 
under investigation and the characteristics of the laser light that is used, the 
extent of spatial alignment that occurs. 
Specific insight is also needed  on the relative importance of angle dependent 
ionization on the one hand and molecular reorientation on the other in making 
sense of measured anisotropies in fragment ion distributions. In this Letter we 
present results of a comprehensive study that enables predictions to be made of 
the propensity of molecules to spatially align in intense, pulsed, polarized light on 
the basis of parameters that are readily accessible. We show that
existing experimental data, spanning work done over the past decade by several groups
 including our own is
explained by our model. We also show that it is possible to 
align molecules even with extremely short light pulses irrespective of the 
polarizability. 

In any given analysis of spatial alignment three factors play a crucial role: 
(i) the peak intensity of the laser pulse, (ii) its temporal duration, and (iii) 
the ratio of the molecular polarizability (ground or excited state) to the 
moment of inertia ($R={\frac {\rm \alpha}{\rm I}}$). At high enough intensities, 
molecular hyperpolarizabilites may also be significant although their role in the 
alignment dynamics has not hitherto been explicitly considered. It is also well 
established that field ionization of molecules is ubiquitous with short pulse 
lasers. The important conclusion of the field ionization model relevant to 
alignment is the breakup of the molecule at a critical distance R$_{c}$ that 
is larger than the equilibrium internuclear separation R$_{e}$ \cite{rc}. The 
stretching of the internuclear axis increases the moment of inertia and leads to 
a slowing down of reorientation; dissociation at R$_{c}$ implies that the 
molecule will dissociate before the peak intensity is reached, except for 
ultrashort light pulses. 

The process of alignment is modeled by considering a rotor in a time 
dependent $\bf{E}$ field \cite{Dion}. In this case the interaction Hamiltonian 
is given by H$_{I}$ = -$\bf{\mu}$$.$$\bf{E}$ where 
$\bf{\mu}=\bf{\mu}_{0}+{\frac {\rm 1}{\rm 2}}\bf{\alpha}\bf{E}+{\frac {\rm 1}{\rm 24}}
\bf{\gamma}\bf{E}\bf{E}$. The Lagrangian is: 
\begin{equation}
L={\frac {\rm I}{\rm 2}}{{\biggl[}{\dot\theta}^2+{sin^2}{\theta}
{\dot\phi}^2{\biggr]}}+{\bf{\mu}}.{\bf{E}}.
\end{equation}
The equation of motion is then given by
\begin{equation}
{\frac {\rm d^2\theta}{\rm dt^2}}=-{\frac {\rm \alpha_{eff}}{\rm 2I}{\bf{E}}(t)^{2}
sin2\theta-2{\frac {\rm \dot r}{\rm r}\dot\theta}},
\end{equation}
where $\alpha_{eff}$ is the effective polarizability. In the low-field limit this is 
the linear polarizability. The modification of this in high fields is discussed later. 
To keep our calculations as realistic as possible we have used a Gaussian laser 
pulse, $E(t)=E_{0}e^{-\frac {\rm t^{2}}{\rm 2\tau^{2}}}cos{\omega}t$. 
Spatial variations within the laser beam are not taken into 
account since it has been shown recently that  
intensity selective experiments minimize focal volume effects \cite{spatial}.
 In Eq. (1), the first term causes reorientation
while the second term, the so-called damping term, impedes the 
motion of the molecular 
axis towards the light field vector. In the field ionization Coulomb explosion model 
\cite{ljf1}, the damping term arises due to the elongation of the molecular axis 
from $R_{e}$ to $R_{c}$ after the removal of one or two electrons by 
tunnel/over the barrier ionization. 

We present our results as follows. We first consider the simplest case 
where the polarizability is linear and no damping term is present. It is shown 
that even such a picture is of wide-ranging validity and utility. We then 
discuss results  of calculations that take into account both damping and 
the hyperpolarizability, and we examine the extent to which our results 
are modified. Our calculations are compared with experimental observations. 

Eq. (2) is solved used a 4th-order Runge-Kutta algorithm for a range 
of light field parameters. We consider the laser intensity range 
10$^{12}$ - 10$^{15}$ W cm$^{-2}$, and pulse durations from 
40 fs to 2 ps. In the linear case, molecule-specific calculations have 
not been done. Instead the $A$-parameter has been taken to lie in the 
interval 2$\times$10$^{4}$ - 6$\times$10$^{7}$ (covering a wide 
range of light and heavy species, amongst them H$_{2}$, N$_{2}$, 
CS$_{2}$ and I$_{2}$). The initial direction of the molecular axis, 
$\theta_{0}$, is taken to be random in space. After the light pulse is 
switched on, the angular position is calculated as a function of time for 
various values of $\theta_{0}$ as shown in figure 1(a). From this one 
obtains a plot of $\theta_{f}$ vs. $\theta_{i}$ where $\theta_{i}$ is 
the initial angular position and $\theta_{f}$ is the angular orientation 
of the molecular axis at a particular instant. This is shown in figure 1(b). 
It is easy to show that the angular distribution of the molecular axis is 
proportional to ${({\frac {\rm d\theta_{f}}{\rm d\theta_{i}})}}^{-1}$. 
This procedure can no longer be used when the applied field is so strong 
that the molecular axis crosses $\theta$=0. In such cases we have 
used a counting method by interpolating the relation between 
$\theta_{f}$ and $\theta_{i}$ to obtain the angular distribution.
 
The trajectories shown in figure 1 (a) correspond to H$_2$ exposed to 
20 fs pulses at 10$^{15}$ W cm$^{-2}$. As can be seen from the slope 
in figure 1(b) the extent of reorientation in H$_2$ is negligible under these 
conditions. Similar calculations were carried out for a range of parameters 
as specified above, and results are shown in figure 2. The surface 
demarcates the regions where spatial alignment is significant from those 
where no significant reorientation occurs. All points {\em above} the surface, 
i.e. in regions of larger intensity, polarizability and pulse duration  
correspond to molecules that are significantly aligned,  while the opposite 
holds for points that lie below the surface. Note that we are dealing here
only with linear polarizabilities. The nonlinear polarizability components
serve only to strengthen the alignment. Thus, the demarcation based on 
$\alpha$ alone is very rigorous.

It is to be noted that the above calculations pertain to the position of
the molecular axis. 
However alignment is deduced from the anisotropy of fragment ions. To make 
the connection with experimental data it should be noted that the angular 
distributions that are shown are those that would practically be measured 
using a spectrometer with a small acceptance angle. Thus, our calculations 
are specially relevant to the angular distributions of highly charged ions 
(that possess large kinetic energies). This is also important in the context 
of the residual angular momentum as the molecule rotates. This is 
sufficiently large to cause significant rotation on the time scale of the 
laser pulse but negligible compared to the energy of the fragment ions 
which are typically in excess of 1 eV. Illustrative experimental data for 
various molecules are 
also shown in the figure and we note the excellent agreement that is 
obtained between our model and measured data. 

We now consider the damping term. Ignoring this term essentially assumes 
that the molecule is a rigid rotor in the intense field. However, as already 
noted, when the field is sufficiently large, the ionization dynamics occur 
through an enhanced ionization (EI) mechanism wherein one or two 
electrons are removed at the first ionization step. Subsequent to the first 
ionization step that occurs at the equilibrium internuclear separation, the 
two residual atomic ions mutually repel each other, leading to an increase 
in the bond length. This results in one or more Stark-shifted electronic 
levels rising above the potential barrier that separates the atomic cores, at 
which point multiple electron ejection occurs, leading to molecular 
fragmentation. EI can modify the reorientation rate in 
two ways. Firstly, as the moment of inertia increases, the magnitude of the 
first term in Eq. (1) will reduce. In addition, the damping term will come 
into play, leading to a further decrease in the rate at which the molecule 
rotates towards the light field vector. It is important to investigate 
the extent EI might modify our first-order calculations. 
Since EI parameters are available only for a few molecules, we have 
carried out these calculations for some standard cases. These can be extended 
to any other molecule once the relevant parameters are known either by 
calculation or experiment. 

Fig. 3 (a) shows the angular distribution for H$_{2}$ for a pulse duration 
of 40 fs at a peak laser intensity of 10$^{15}$ W cm$^{-2}$, with and 
without the damping term. It is clear that the reorientation of H$_{2}$ is 
not  significantly affected when the damping term is included. There are 
two major reasons for this. Firstly, $R$ is extremely large and  the 
torque experienced by H$_2$ is sufficient to induce reorientation despite 
the presence of an opposing force. Secondly, the fact that the ionization 
energy (and hence, the appearance intensity) of H$_2$ is quite high, the 
damping force only comes into play close to the peak of the laser pulse, 
by which time the molecular axis is already aligned with the light polarization 
vector. Interestingly, the width of the angular distribution 
with damping included is actually smaller than when no damping is present. 
This arises due to the fact that the angular velocity without damping is larger, 
causing the molecular axis to execute large amplitude oscillations about 
$\theta$=0; hence, there will exist instants at which the peak of the 
angular distribution will shift away from zero. 

A contrary situation is depicted in Fig. 3(b) when linear CS$_2$ molecules 
are exposed to 100 fs light fields. In this case, the lower ionization energy 
of the molecule coupled with the relatively small value of $A$, leads to a 
situation where there is virtually no reorientation of the S-C-S axes with the 
direction of the ${\bf E}$ field at the point at which dissociation occurs. 
Strong alignment can be expected if it is assumed that the molecule survives 
undissociated till the peak of the laser pulse, a fact contrary to experimental 
observation, and illustrates the essentiality of EI in any such model. 
Similar calculations have been carried out for other molecules like 
N$_{2}$ and I$_{2}$. The situation for N$_{2}$ is similar to H$_{2}$ 
because of the similarity in the relevant parameters. In the case of I$_{2}$, 
reorientation of the molecule is not significant even without the damping 
term. Once the damping term is included there is only a small deflection 
of the molecular axis. 

Hitherto, only the polarization response that arises from the linear term 
has been considered. To what extent is this justified, especially at 
intensities in the range of 
10$^{12}$ - 10$^{15}$  W cm$^{-2}$? It is important to note that 
hyperpolarizabilites are significant only at the highest intensities. This is 
obvious when we compare  the integrals which define the work done by the 
field on the molecule by each order of the hyperpolarizability. It can be 
expected that  for longer pulses, a model based on linear polarizability is 
sufficient since the dissociation of the molecule occurs on the rising edge 
of the pulse. However, as the pulse duration becomes shorter ($<$50 fs), 
the molecule will survive till the maximum intensity is reached, and the 
reorientation due to the higher order terms will become comparable to that 
due to the linear term, and may even exceed it! 

To incorporate hyperpolarizability in our calculations, certain 
approximations need to be made. Firstly, the magnitude of the second- 
and higher-order susceptibilities is not known in most cases \cite{wang}. Moreover, 
these quantities are tensors with numerous components and it is difficult 
to consider all the components in an exact way. We consider here the 
case for H$_2$ taking into account the third order term due to the 
electronic response ${\gamma}_{e}$. The equation of motion is 
modified as follows: $\alpha$$sin2\theta$$E^{2}$$\rightarrow$
$\alpha$$sin2\theta$$E^2$+${\gamma}_{e}$$sin4{\theta}$$E^4$
+$\ldots$.  

Fig. 4 shows the effect of nonlinearity on the reorientation of the 
molecular axis for H$_2$ acted on by a 20 fs pulse.The inclusion of 
only the third order term \cite{wang} leads to a significantly larger reorientation 
of the H-H axis as compared to the case when only the linear term is 
considered. As noted previously, such effects will be significant only for 
very short pulses. This is shown in figure 3(a) for the case of a 
40 fs pulse wherein the effect of the third order term is much smaller than 
that for the 20 fs case. We have verified that for longer pulses such 
higher order terms need no longer be taken into account. Of course, it is 
obvious that as the laser pulses get shorter even higher order terms will 
begin to play a significant role. One can speculate that heavy molecules, 
like I$_2$, may align with sufficiently short pulses because of the contribution 
from higher order terms. It is clearly necessary to test this conjecture 
experimentally since very little is known about the high order polarizabilities 
of almost all molecules.

In summary, we have considered molecular reorientation using a classical 
model. The justification for using a classical model is two-fold. In intense 
field-molecule interactions, classical analysis have been shown to be extremely 
fruitful in explaining much of the experimental data. Also, exact 
time-dependent analysis of the molecular response to a high-intensity pulsed 
light field is presently not feasible. We show that, 
despite the obvious limitations of classical models, the results are considerable 
utility in understanding a large body of experimental work on alignment, and for
enabling predictions to be made on whether or not, and under what
circumstances, molecules will be spatially aligned when subjected to intense,
polarized, short-duration light fields. The rigor of the model is demonstrated. 
We have also incorporated, for the first time, 
(i) the role of enhanced ionization in the reorientation of molecules and (ii) 
the role of hyperpolarizability. It is shown that {\em higher order 
contributions to the dipole moment are very important for extremely short 
light pulses}. It is predicted that this will lead to alignment of molecules 
even for sub-50 fs pulses that are becoming increasingly accessible to
experimentalists.

The TIFR high energy, femtosecond laser facility has been set up with 
substantial funding from the Department of Science and Technology, Government of India.

\begin{figure}
\caption{(a) Time evolution of the molecular alignment for various initial 
orientations at a peak intensity of 10$^{15}$ W cm$^{-2}$ and temporal width 
of 20 fs. (b) Alignment of the molecular axis at the point of breakup predicted 
by the EI model as a function of initial orientation. }
\end{figure}

\begin{figure}
\caption{(a) Alignment of molecules for various conditions of peak intensity, 
pulse duration and R (see text). Points lying below the surface correspond to 
the case of no alignment while all points lying on the surface, and above it, 
lead to the molecular axis being aligned along the light polarization vector. 
(b) Experimental data for some typical molecules: CO [1], CO$_2$ [1], H$_2$,
N$_2$ [3], I$_2$ [3], CS$_2$ (picosecond data [2], femtosecond 
data [11]). The axes ranges are the same as in (a).}
\end{figure}

\begin{figure} 
\caption {Alignment dynamics calculated with enhanced ionization taken 
into account (a) H$_2$ (b) CS$_2$. L$\equiv$linear polarizability, 
ND$\equiv$no damping, D$\equiv$damped rotation, NL3$\equiv$linear and 
third order polarizability.}
\end{figure}

\begin{figure}
\caption{Alignment of molecular axes for H$_{2}$ assuming nonlinear 
contributions to the polarizability at a peak intensity of 10$^{15}$ 
W cm$^{-2}$ and temporal duration of 20 fs. The symbols are explained in fig.3}
\end{figure}

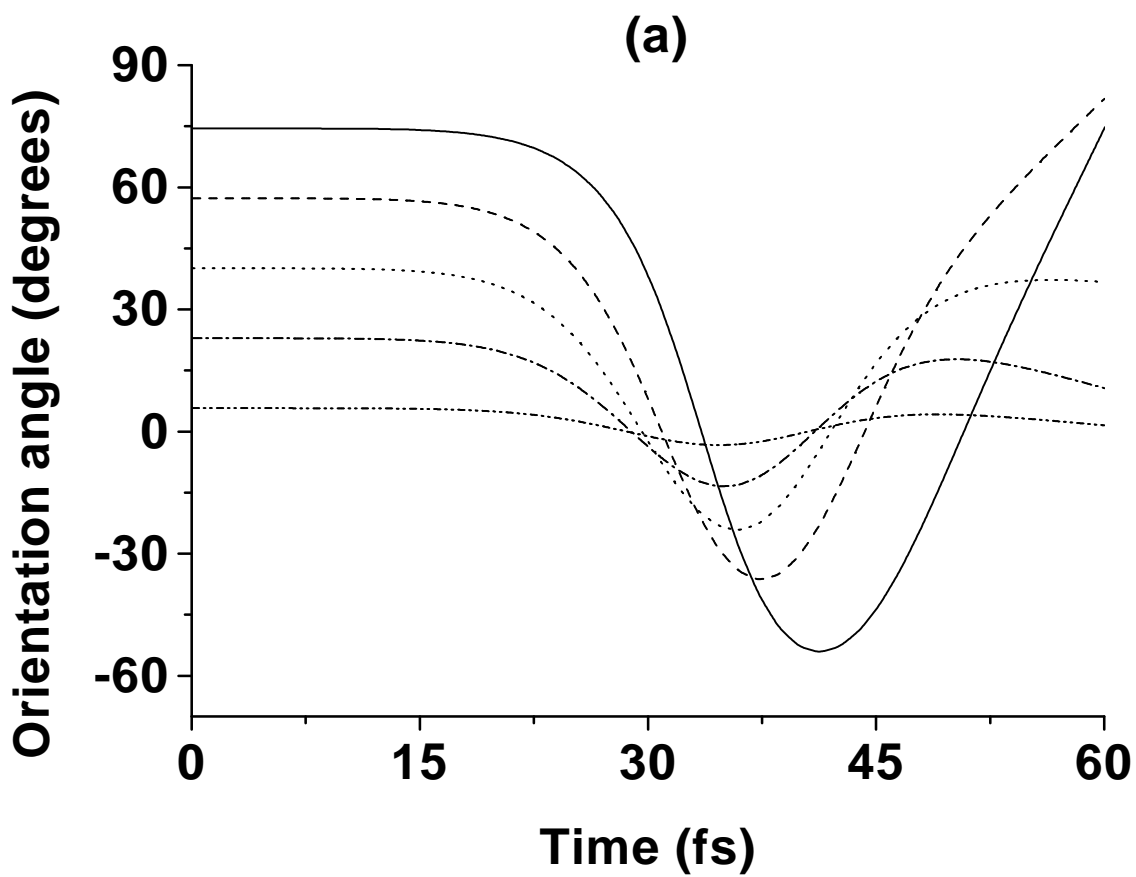
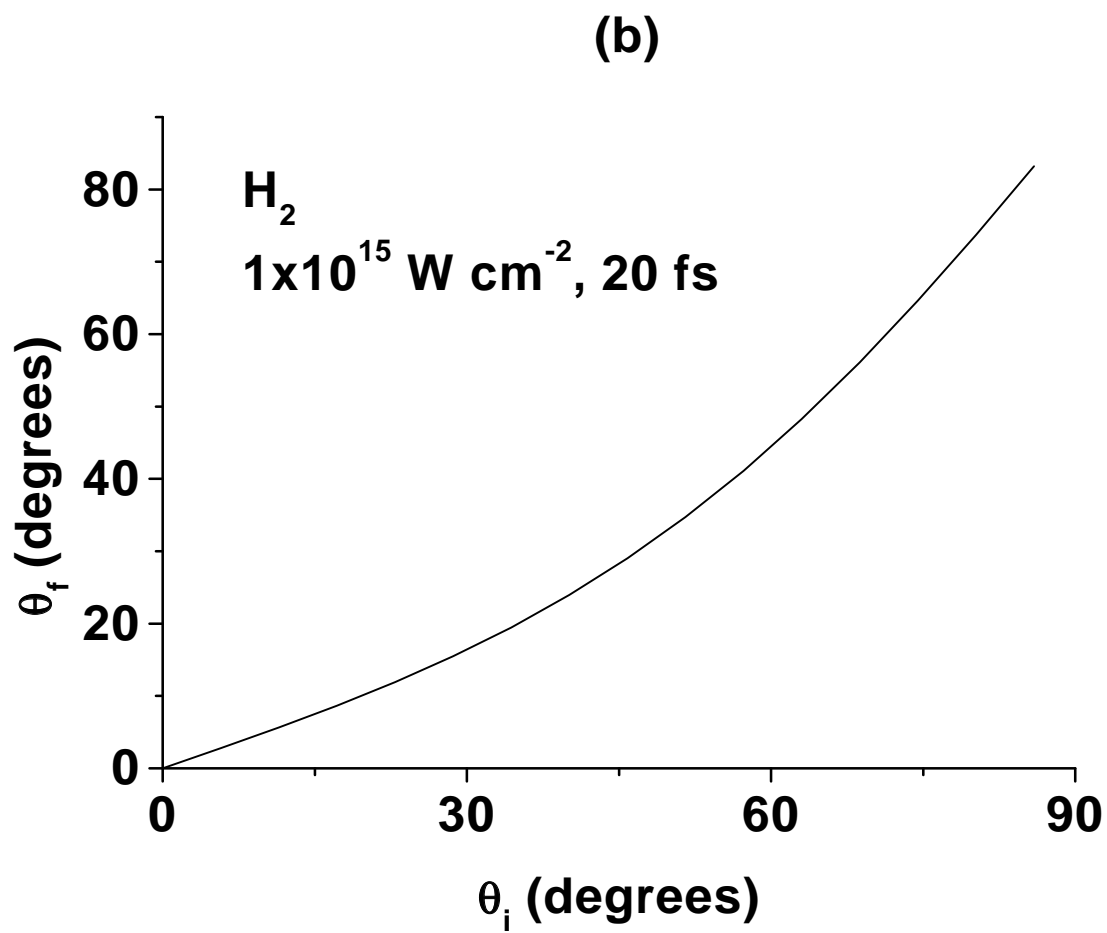
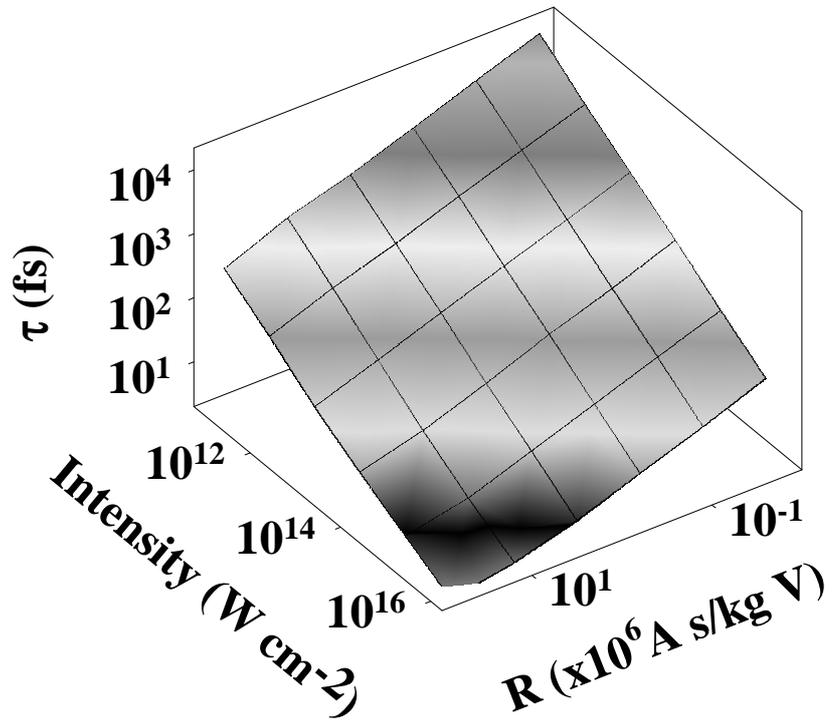
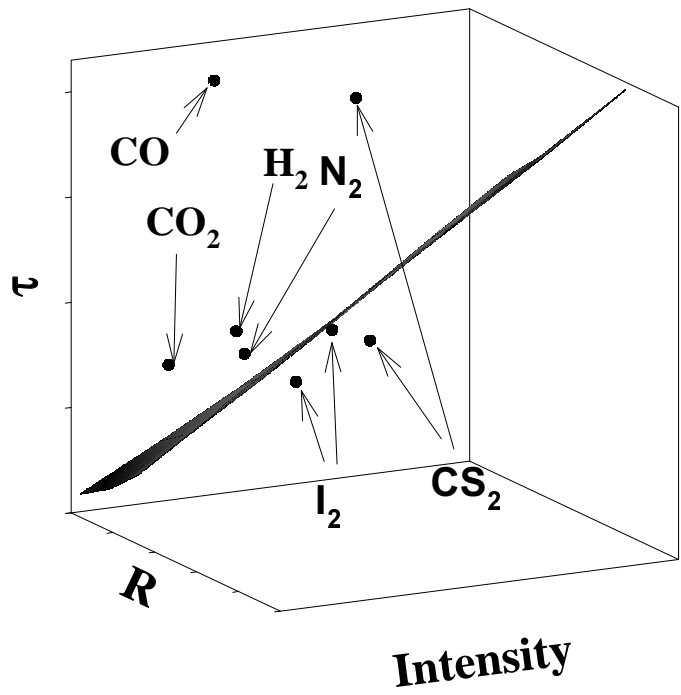
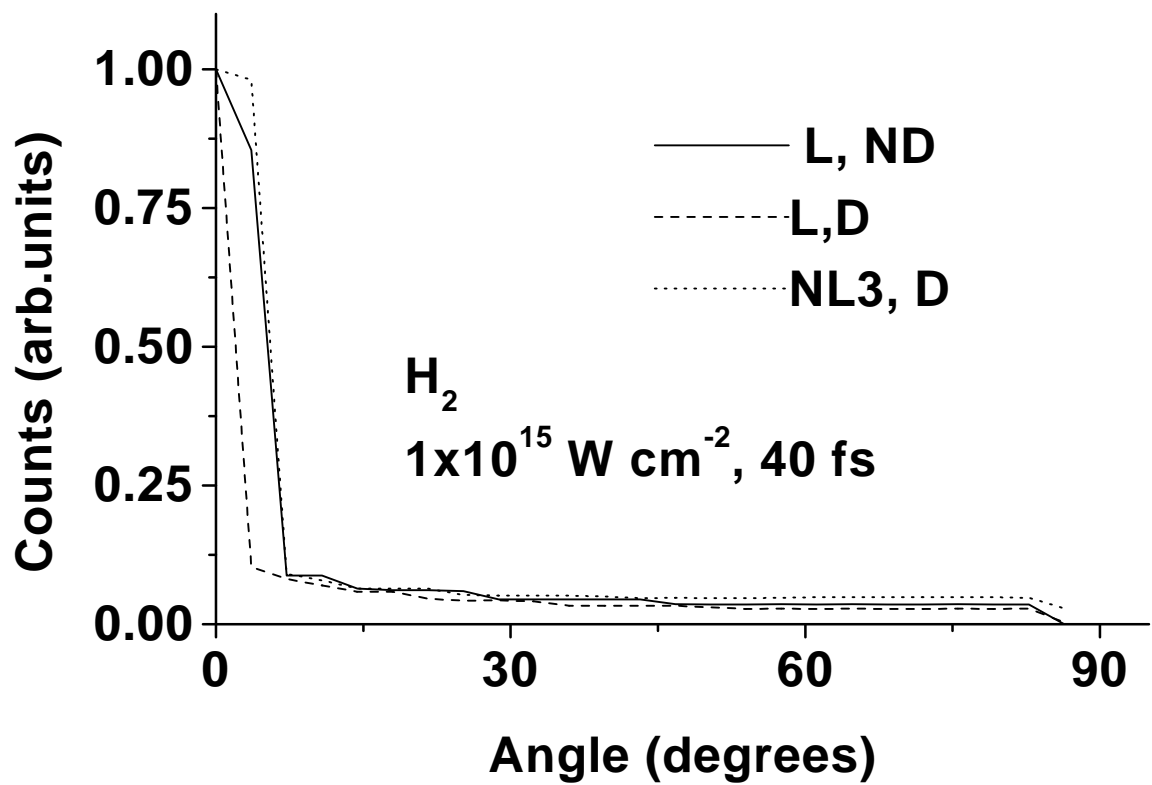
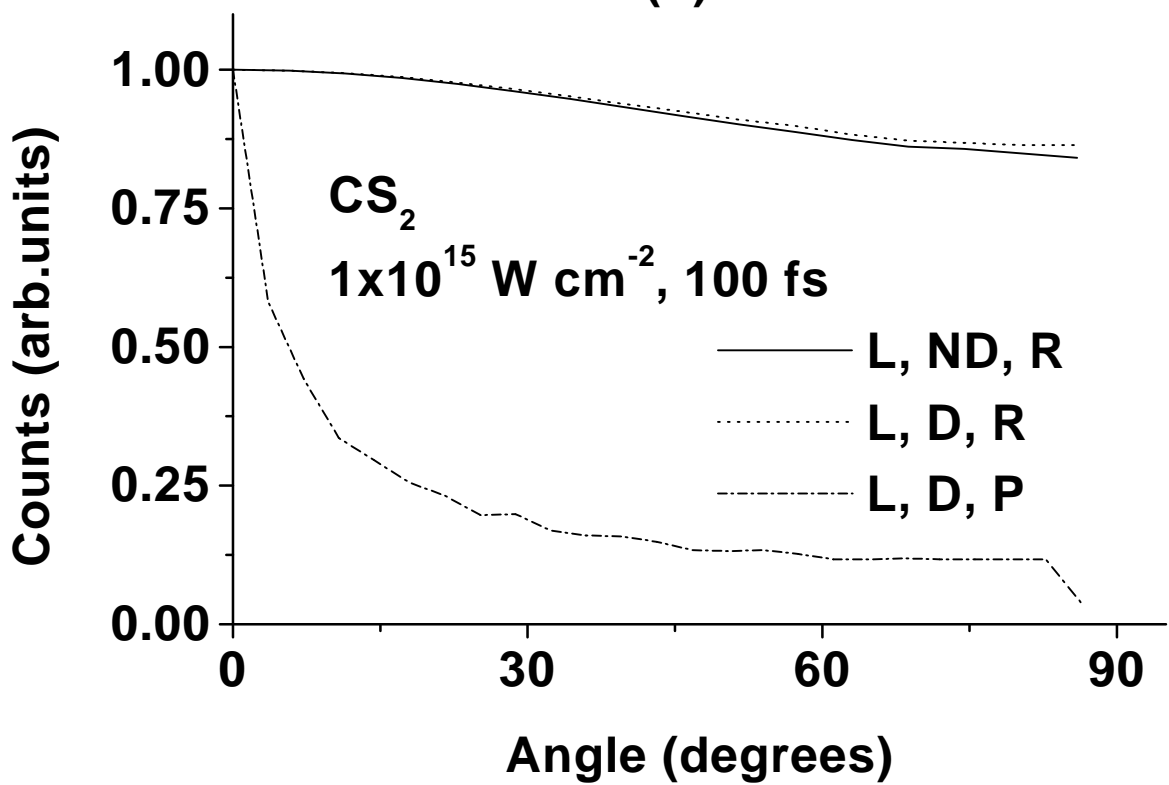
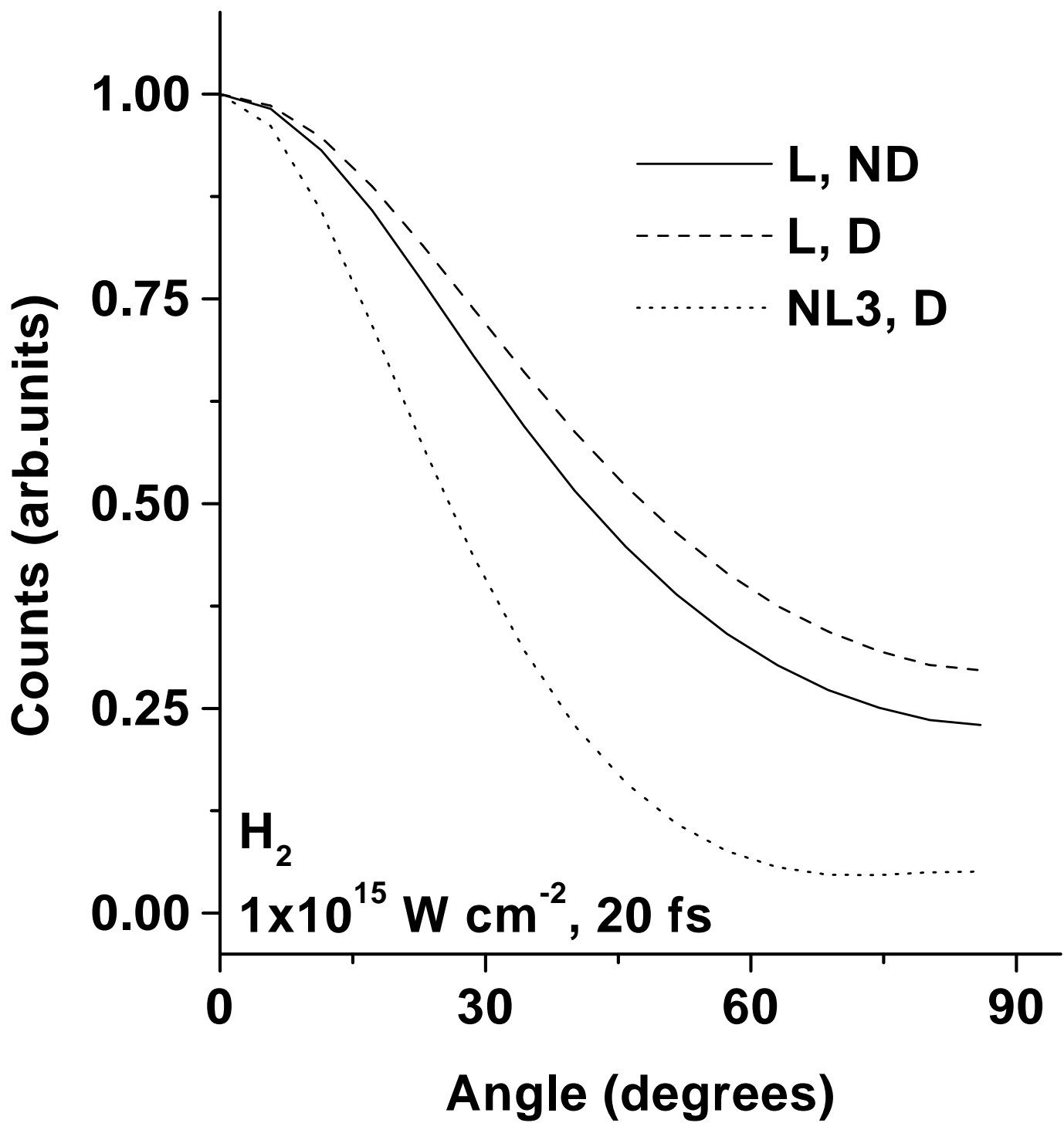
\end{document}